\documentclass[prd,showpacs, twocolumn,amsmath,amssymb]{revtex4}

\usepackage{amssymb}
\usepackage{graphicx}
\usepackage{dcolumn}
\usepackage{bm}
\usepackage{epsfig}
\usepackage{amsfonts}
\usepackage{keyval,graphicx}
\usepackage{textcomp,wasysym}
\usepackage{subfigure}
\usepackage[none]{hyphenat}
\usepackage{xcolor}

\begin{document}
\newcommand{\chisq}[1]{$\chi^{2}_{#1}$}
\newcommand{\etap}{\eta^{\prime}}
\newcommand{\pip}{\pi^{+}}
\newcommand{\pim}{\pi^{-}}
\newcommand{\gam}{\gamma}
\newcommand{\piz}{\pi^{0}}
\newcommand{\rhoz}{\rho^{0}}
\newcommand{\az}{a_{0}(980)}
\newcommand{\fz}{f_{0}(980)}
\newcommand{\pipm}{\pi^{\pm}}
\newcommand{\psip}{\psi^{\prime}}
\newcommand{\psipp}{\psi^{\prime\prime}}
\newcommand{\jpsi}{J/\psi}
\newcommand{\ar}{\rightarrow}
\newcommand{\GeV}{GeV/$c^2$}
\newcommand{\MeV}{MeV/$c^2$}
\newcommand{\br}[1]{\mathcal{B}(#1)}
\newcommand{\cinst}[2]{$^{\mathrm{#1}}$~#2\par}
\newcommand{\crefi}[1]{$^{\mathrm{#1}}$}
\newcommand{\crefii}[2]{$^{\mathrm{#1,#2}}$}
\newcommand{\crefiii}[3]{$^{\mathrm{#1,#2,#3}}$}
\newcommand{\HRule}{\rule{0.5\linewidth}{0.5mm}}


\title{\Large \boldmath \bf First observation of $\eta(1405)$ decays into $f_{0}(980)\pi^0$}

\author{\small
M.~Ablikim$^{1}$, M.~N.~Achasov$^{5}$, D.~Alberto$^{41}$,
D.J.~Ambrose$^{38}$, F.~F.~An$^{1}$, Q.~An$^{39}$, Z.~H.~An$^{1}$,
J.~Z.~Bai$^{1}$, R.~B.~F.~Baldini Ferroli$^{17}$, Y.~Ban$^{25}$
% ...
\\
\vspace{0.2cm}
(BESIII Collaboration)\\
\vspace{0.2cm} {\it
$^{1}$ Institute of High Energy Physics, Beijing 100049, P. R. China\\
% ...
$^{5}$ G.I. Budker Institute of Nuclear Physics SB RAS (BINP), Novosibirsk 630090, Russia\\
% ...
$^{17}$ INFN Laboratori Nazionali di Frascati , Frascati, Italy\\
% ...
$^{25}$ Peking University, Beijing 100871, P. R. China\\
% ...
$^{38}$ University of Rochester, Rochester, New York 14627, USA\\
$^{39}$ University of Science and Technology of China, Hefei 230026, P. R. China\\
% ...
$^{41}$ University of Turin and INFN, Turin, Italy\\
% ...
}}

\vspace{0.4cm}



\collaboration{BESIII Collaboration}


\begin{abstract}
The decays $J/\psi \rightarrow \gamma \pi^+\pi^-\pi^0$ and $J/\psi
\rightarrow \gamma \pi^0\pi^0\pi^0$ are analyzed using a sample of
225~million $\jpsi$ events collected with the BESIII detector. The decay of $\eta(1405)\ar f_{0}(980)\pi^0$ with a large isospin violation is observed for the first time. The width of the $f_{0}(980)$ observed in the dipion mass spectra is anomalously narrower than the world average. Decay rates for three-pion decays of the $\eta'$ are
also measured precisely.
\end{abstract}


\maketitle


%
A state near 1440~\MeV~was discovered in $p\overline{p}$ annihilation at rest decaying to $\eta\pi^+\pi^-$ and was subsequently observed decaying to $K\overline{K}\pi$~\cite{ref:first1440}. Considerable theoretical and experimental efforts have since been devoted to understand its nature. It was proposed to be a candidate for a pseudo-scalar glueball~\cite{ref:glue1440,ref:glueFad}; the measured mass, however, is much lower than that obtained from lattice QCD calculations, which is above 2~GeV/c$^2$~\cite{ref:latticeQCD}. Later, experiments produced evidence that this state was really two different pseudo-scalar states, the $\eta(1405)$ and the $\eta(1475)$. The former has large couplings to $a_{0}(980)\pi$ and $K\bar{K}\pi$, while the latter mainly couples to $K^{*}\bar{K}$. A detailed review of the experimental situation can be found in Ref.~\cite{ref:status1440}.

The nature of the well-established light scalars $f_{0}(980)$ and
$a_{0}(980)$ is also a matter of controversy. It is not clear whether
they belong to the light scalar meson nonet or are examples of mesons
beyond the naive quark model (eg. tetra-quark states, hybrids or
$K\bar{K}$
molecules)~\cite{ref:a0f0-1,ref:a0f0-2,ref:a0f0-3,ref:a0f0-4,ref:a0f0-5,ref:a0f0-6}. The
possibility of mixing between the $f_{0}(980)$ and $a_{0}^{0}(980)$
was suggested long ago, and its measurement sheds light on the
nature of these two
resonances~\cite{Achasov:1979xc,Hanhart:2007bd,Achasov:2002hg,
Kerbikov:2000pu,Achasov:2003se,Close:2000ah}.

The three-pion decays of the $\eta'$ have garnered attention because their branching ratios (Br) can
probe isospin breaking~\cite{ref:etaptheo1,ref:etaptheo2}. The ratios of the branching ratios ($r_{\pm}\equiv Br(\eta'\to\pi^+\pi^-\pi^0)/Br(\eta'\to\pi^+\pi^-\eta)$ and $r_{0}\equiv Br(\eta'\to 3\pi^0)/Br(\eta'\to\pi^0\pi^0\eta)$) are related to the strange quark mass and SU(3) breaking~\cite{ref:etaptheo1}.

In this letter, we present the results of a study of
$J/\psi\to\gamma\pi^+\pi^-\pi^0$ and $J/\psi\to\gamma\pi^0\pi^0\pi^0$.
A single structure around 1.4~GeV/$c^2$ in the $\pi^+\pi^-\pi^0$
($\pi^0\pi^0\pi^0$) mass spectrum is observed, associated with a
narrow structure around 980~MeV/$c^2$ in the $\pi^+\pi^-$
($\pi^0\pi^0$) mass spectrum.  This analysis is based on a sample of
$(225.2\pm2.8)\times10^6$ $\jpsi$ events~\cite{ref:jpsitotnumber}
accumulated in the Beijing Spectrometer (BESIII)~\cite{ref:bes3nim}
operating at the Beijing Electron-Positron Collider
(BEPCII)~\cite{bepc2_design}.

BEPCII is a double-ring $e^+e^-$ collider designed to provide
$e^+e^-$ collisions with a peak luminosity of $10^{33}
~\rm{cm}^{-2}\rm{s}^{-1}$ at a beam current of 0.93~A. The cylindrical
core of the BESIII detector consists of a helium-based main drift
chamber (MDC), a plastic scintillator time-of-flight system (TOF), and
a CsI(Tl) electromagnetic calorimeter (EMC), which are all enclosed in
a superconducting solenoidal magnet providing a 1.0~T magnetic
field. The solenoid is supported by an octagonal flux-return yoke with
resistive plate counter muon identifier modules interleaved with
steel. The acceptance of charged particles and photons is 93\% over
4$\pi$ stereo angle, and the charged-particle momentum and photon
energy resolutions at 1~GeV are 0.5\% and 2.5\%, respectively. The BESIII detector is modeled with a Monte Carlo (MC) simulation based on \textsc{geant}{\footnotesize
4}~\cite{ref:geant4,ref:geant4_2}.

The charged-particle tracks in the polar angle range $|\cos\theta|<0.93$
are reconstructed from hits in the MDC. Tracks that extrapolate to
 be within $20~{\rm cm}$ of the
interaction point in the beam direction and  $2~{\rm cm}$ in the
plane perpendicular to the beam are selected.  The TOF and $dE/dx$
information are combined to form particle identification confidence
levels for the $\pi$, $K$, and $p$ hypotheses; each track is
assigned to the particle type that corresponds to the hypothesis
with the highest confidence level.
 Photon candidates are required to have at least $25~{\rm MeV}$ and $50~{\rm MeV}$ of energy in the
EMC regions $|\cos\theta|<0.8$ and  $0.86<|\cos\theta|<0.92$, respectively, and be
separated from all charged tracks by more than $10^\circ$.

For $J/\psi \rightarrow \gamma \pi^+\pi^-\pi^0$, the candidate events are required to have two oppositely charged tracks identified as pions and at least three photon candidates. A four-constraint(4C)
energy-momentum conserving kinematic fit is performed to
the $\gamma\gamma\gamma\pi^+\pi^-$ hypothesis, and $\chi^{2}_{4C}<30$ is required.
For events with more than three photon candidates, the combination with the smallest $\chi^{2}$ is retained. To reject possible background events with two or four photons
in the final state, the 4C-fit probability for an assignment of the $J/\psi
\rightarrow\pi^+\pi^-\gamma\gamma\gamma$ channel must be larger than that of the $J/\psi
\rightarrow\pi^+\pi^-\gamma\gamma$ and the $J/\psi
\rightarrow\pi^+\pi^-\gamma\gamma\gamma\gamma$ channels. The $\pi^0$ candidates are selected by requiring $|M_{\gamma\gamma}-m_{\pi^{0}}|<0.015$~GeV/$c^2$. Events with $|M_{\gamma\pi^{0}}-m_{\omega}|<0.05$~GeV/$c^2$ are rejected to suppress the background from $J/\psi\to\omega\pi^+\pi^-$.

\begin{figure}[htbp]
   \centerline{
   \psfig{file=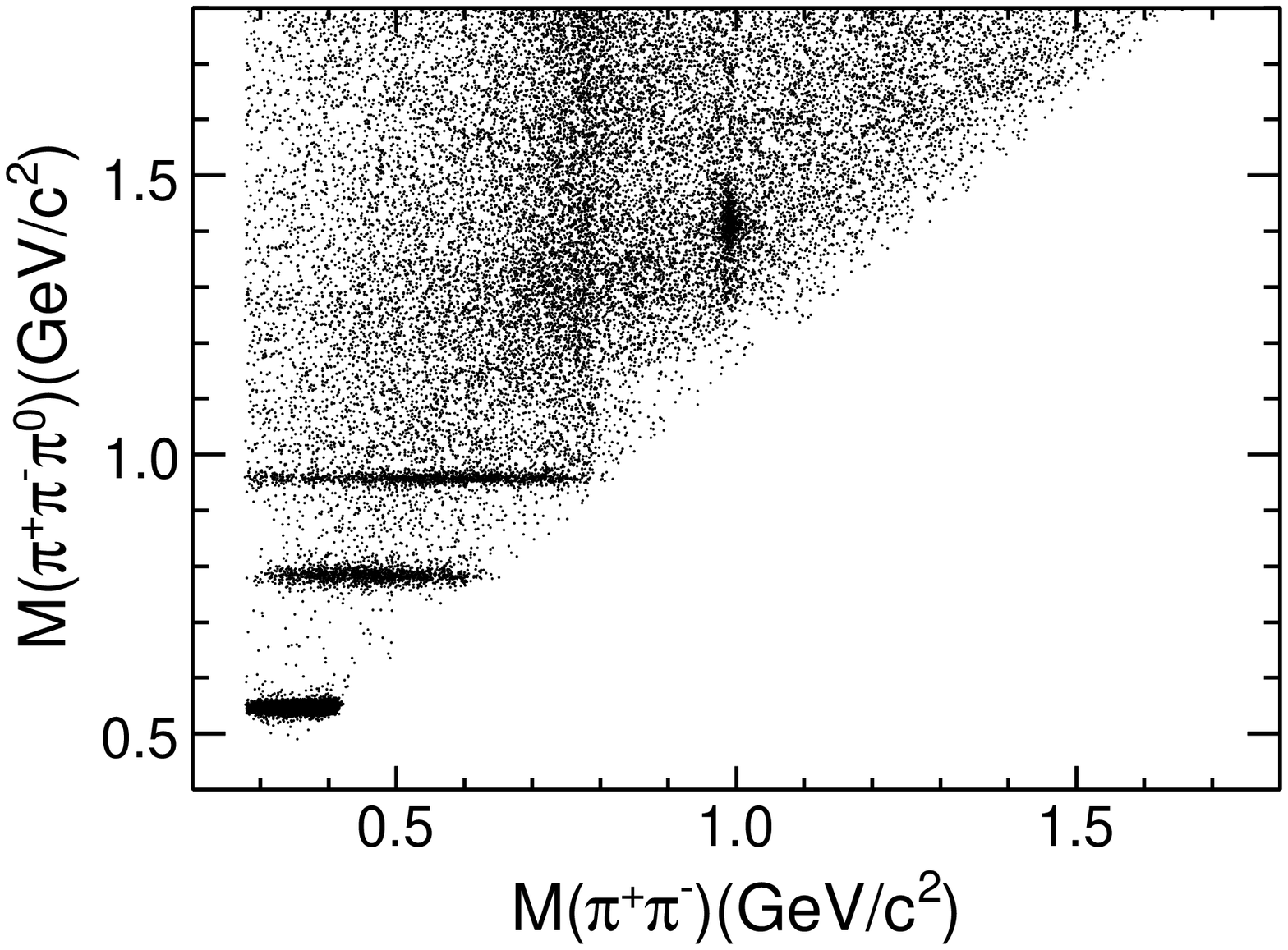,width=4.cm, angle=0}
              \put(-20,30){(a)}
   \psfig{file=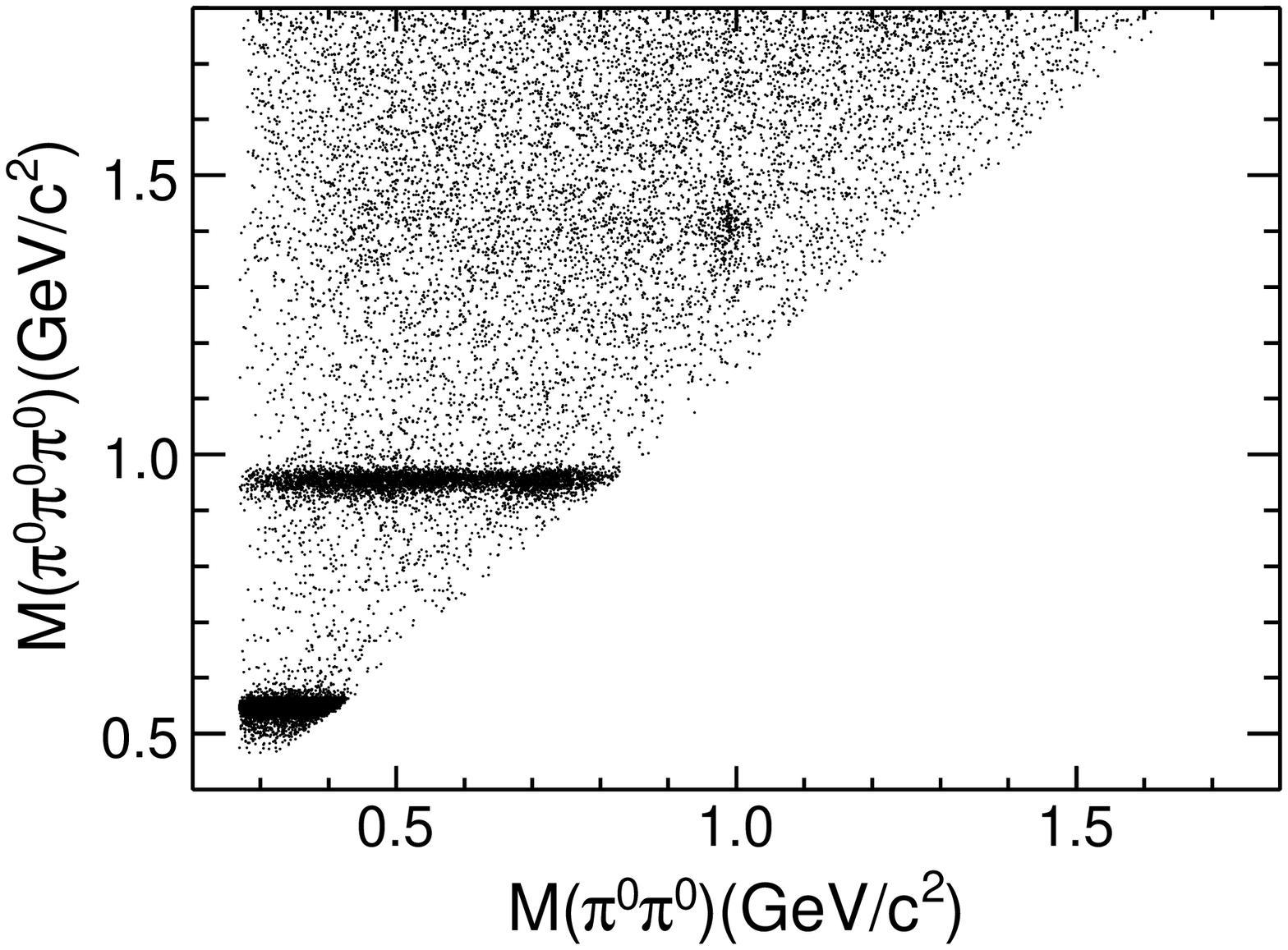,width=4.cm, angle=0}
              \put(-20,30){(b)}}
    \caption{\label{fig:scatter_plot} (a) Scatter plot of $M_{\pi^+\pi^-\pi^0}$ versus $M_{\pi^+\pi^-}$.  (b) Scatter plot of $M_{\pi^0\pi^0\pi^0}$ versus $M_{\pi^0\pi^0}$ (3 entries per event).
    }
\end{figure}

For $J/\psi \rightarrow \gamma 3\pi^0$, the candidate events are
required to have no charged track. The $\pi^0\rightarrow\gamma\gamma$
candidates are formed from pairs of photon candidates that are
kinematically fit to the $\pi^0$ mass, and the $\chi^{2}$ from the
kinematic fit with 1~degree of freedom are required to be less than
25. True $\pi^0$ mesons decay isotropically in the $\pi^{0}$ rest
frame, and their decay distributions are flat, contrary to $\pi^0$
candidates originating from wrong photon combinations. To remove wrong
photon combinations, the decay angle, defined as the polar angle of a
photon in the $\pi^{0}$ rest frame, is required to satisfy
$|\cos\theta_{decay}|<0.95$.  Events with at least seven and less than
nine photons, which form at least three distinct $\pi^0$ candidates,
are selected. A 7C kinematic fit is performed to the $J/\psi
\rightarrow \gamma 3\pi^0$ hypothesis (constraints are the 4-momentum
of $J/\psi$ and the three $\pi^0$ masses), and $\chi^{2}_{7C}<60$ is
required. If there is more than one combination, the combination with
the smallest $\chi^{2}_{7C}$ is retained. Events with
$|M_{\gamma\pi^{0}}-m_{\omega}|<0.05$~GeV/$c^2$ are rejected to
suppress the background from J$/\psi\to\omega\pi^0\pi^0$.

The distributions of the selected events in the
$M_{\pi^{+}\pi^{-}\pi^{0}}$-$M_{\pi^{+}\pi^{-}}$ and
$M_{\pi^{0}\pi^{0}\pi^{0}}$-$M_{\pi^{0}\pi^{0}}$ planes are shown in
Fig.~\ref{fig:scatter_plot} (a) and (b), respectively. The clusters
corresponding to $\eta\rightarrow 3\pi$, $\eta^{\prime}\rightarrow
3\pi$ and $\eta(1405)\rightarrow f_0(980)\pi^{0}$ can be clearly
discerned; the $\omega$ signal is also evident in
Fig.~\ref{fig:scatter_plot} (a), which mainly comes from the background
channel $J/\psi\rightarrow \omega\pi^0$, while it is not observed in
the neutral channel, as expected from charge conjugation symmetry.

To confirm that the apparent signal for $\eta(1405)\rightarrow f_0(980)\pi^{0}$ is not caused by background events, we perform a study with an inclusive MC sample of $2.25\times10^8$ $J/\psi$ events generated according to the Lund-Charm model~\cite{ref:bes3gen} and the Particle Data Group (PDG) decay tables~\cite{PDG}. After the same event selection as above, neither $\eta(1405)$ nor $f_0(980)$ are seen in the mass spectra. Non-$f_0(980)$ or non-$\eta(1405)$ processes are studied using the $f_0(980)$ sidebands (0.88~GeV/$c^2<M_{\pi^{+}\pi^{-}(\pi^{0}\pi^{0})}<0.93$~GeV/$c^2$ and
1.05~GeV/$c^2<M_{\pi^{+}\pi^{-}(\pi^{0}\pi^{0})}<1.10$~GeV/$c^2$) or the $\eta(1405)$ sidebands (1.15~GeV/$c^2<M_{\pi^+\pi^-\pi^0(\pi^{0}\pi^{0}\pi^{0})}<1.25$~GeV/$c^2$ and 1.55~GeV/$c^2<M_{\pi^+\pi^-\pi^0(\pi^{0}\pi^{0}\pi^{0})}<1.65$~GeV/$c^2$). No peaking structures are observed.

\begin{figure}[htbp]
   \centerline{
   \psfig{file=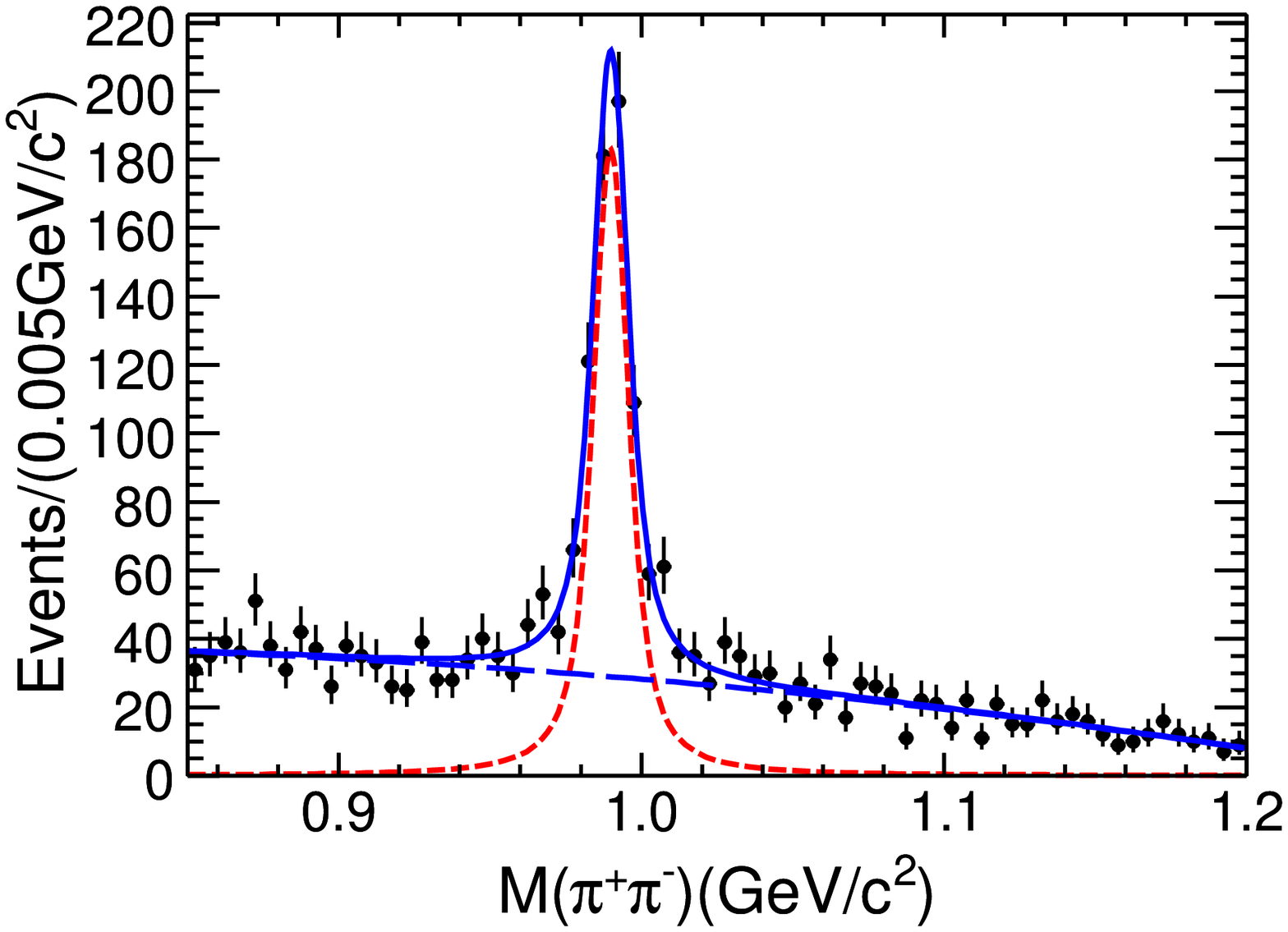,width=4.cm, angle=0}
              \put(-25,65){(a)}
   \psfig{file=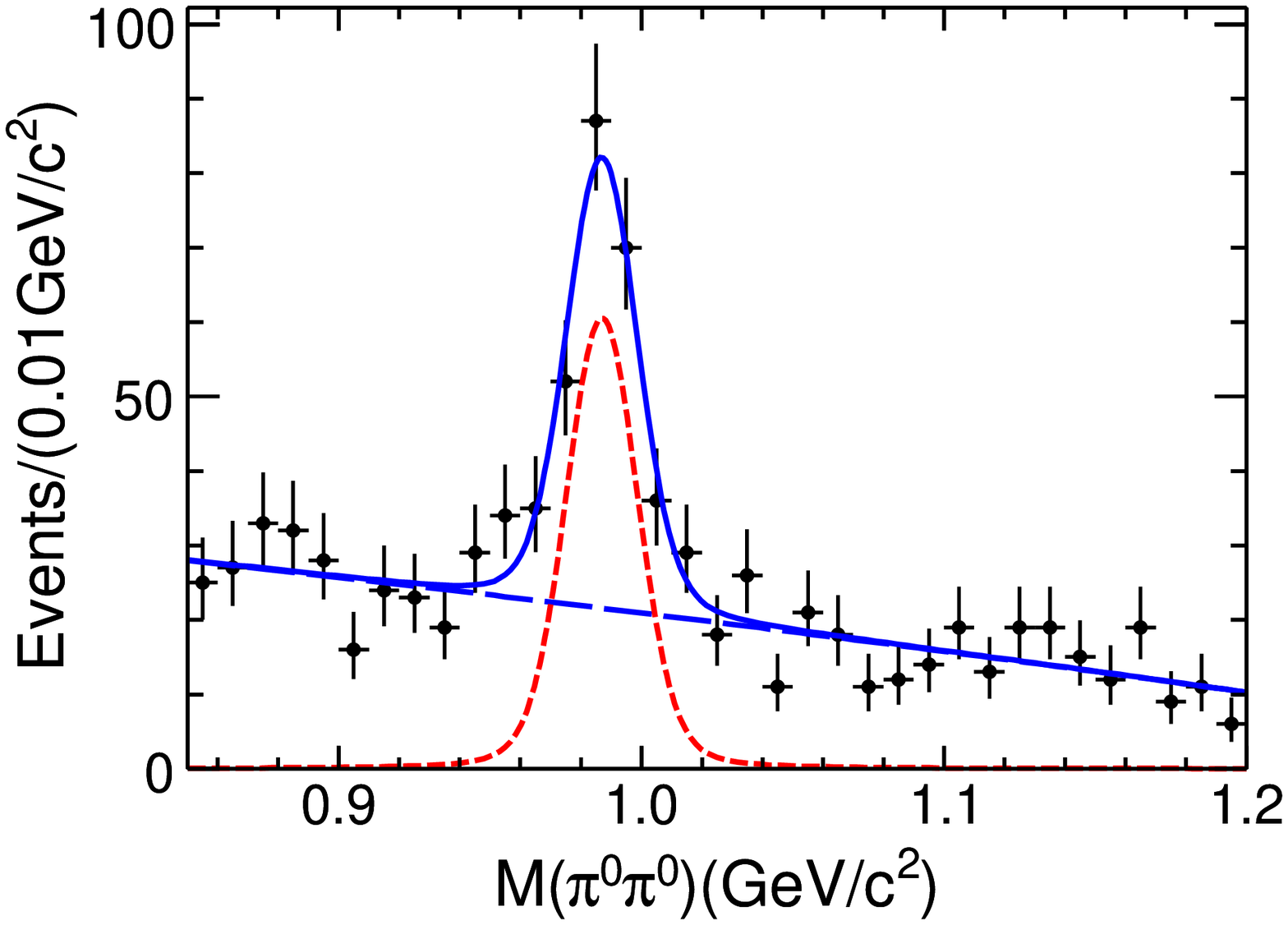,width=4.cm, angle=0}
              \put(-25,65){(b)}}
    \caption{\label{fig:fit_result_2pi}The $\pi^+\pi^-$ and $\pi^{0}\pi^{0}$ invariant mass
spectra with $\pi^+\pi^-\pi^0(3\pi^0)$ in the $\eta(1405)$ mass
region. The solid curve is the result of the fit described in the
text. The dotted curve is the $f_{0}(980)$ signal. The dashed curve
denotes the background polynomial.}
\end{figure}

Fig.~\ref{fig:fit_result_2pi} (a) and (b) show the $\pi^+\pi^-$ and
$\pi^0\pi^0$ mass spectra with the requirement
1.3~GeV/c$^2<M_{\pi^+\pi^-\pi^0(3\pi^0)}<1.5$~GeV/c$^2$. A fit is
performed to the $\pi^{+}\pi^{-}$ mass spectrum with the $f_0(980)$
signal parameterized by a Breit-Wigner function convolved with a
Gaussian mass resolution function plus a second-order Chebychev
polynomial background function.  The mass, width and number of events
of the $f_0(980)$ obtained from the fit are $m =
989.9\pm0.4$~MeV/$c^2$, $\Gamma = 9.5\pm 1.1$~MeV/c$^2$ and $N =
706\pm41$, respectively. A fit to the $\pi^{0}\pi^{0}$ mass spectrum,
shown in Fig.~\ref{fig:fit_result_2pi} (b), is performed in a similar
fashion. The mass, width and number of events of the $f_0(980)$
obtained from the fit are $m = 987.0\pm1.4$ MeV/$c^2$, $\Gamma =
4.6\pm 5.1$ MeV/c$^2$(less than 11.8\MeV~at 90\% C.L.) and $N = 190\pm30$, respectively. The measured
width of the $f_{0}(980)$ is much narrower than the world average.

\begin{figure}[htbp]
\centerline{
   \psfig{file=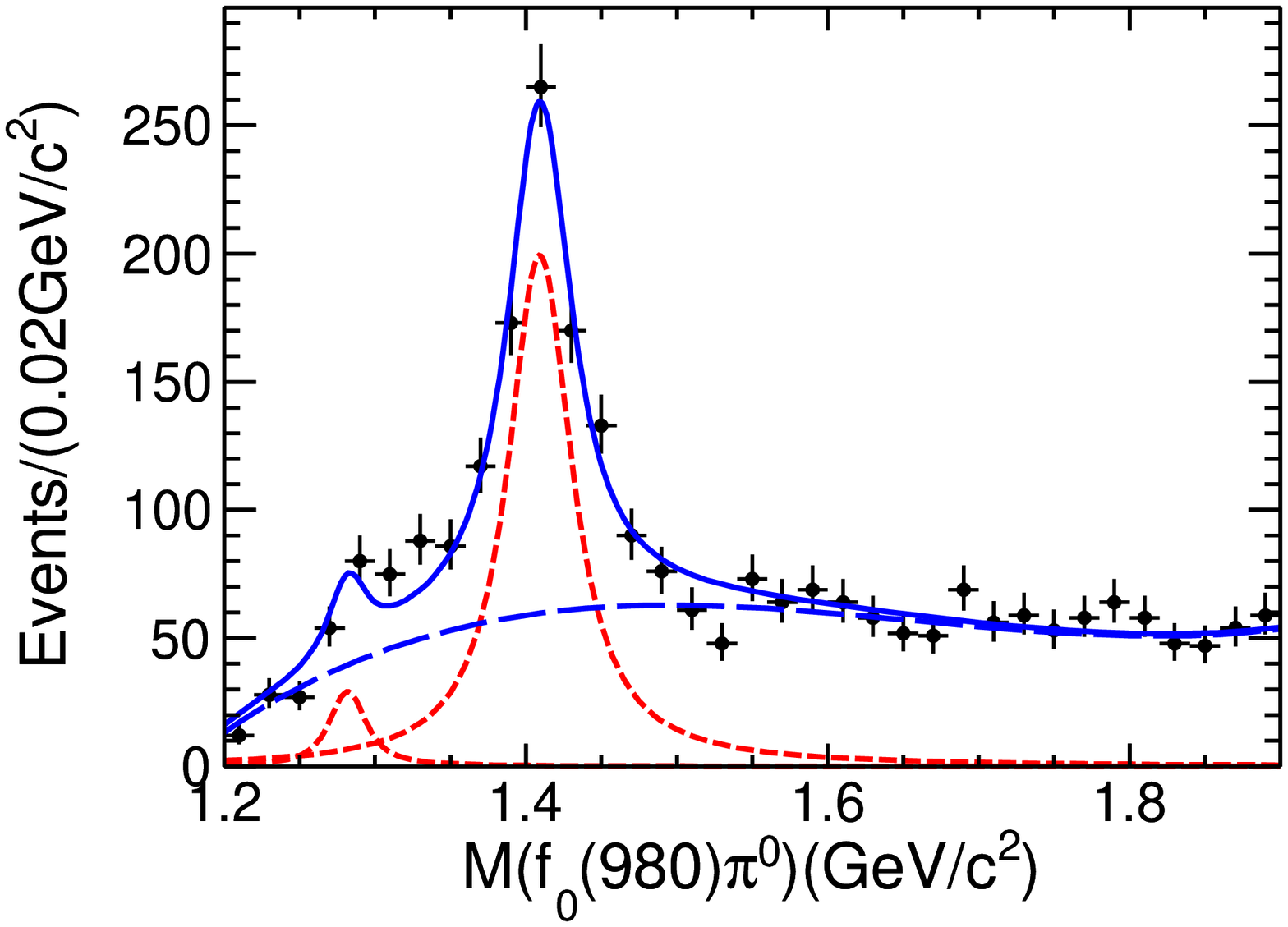,width=4.cm, angle=0}
              \put(-20,70){(a)}
   \psfig{file=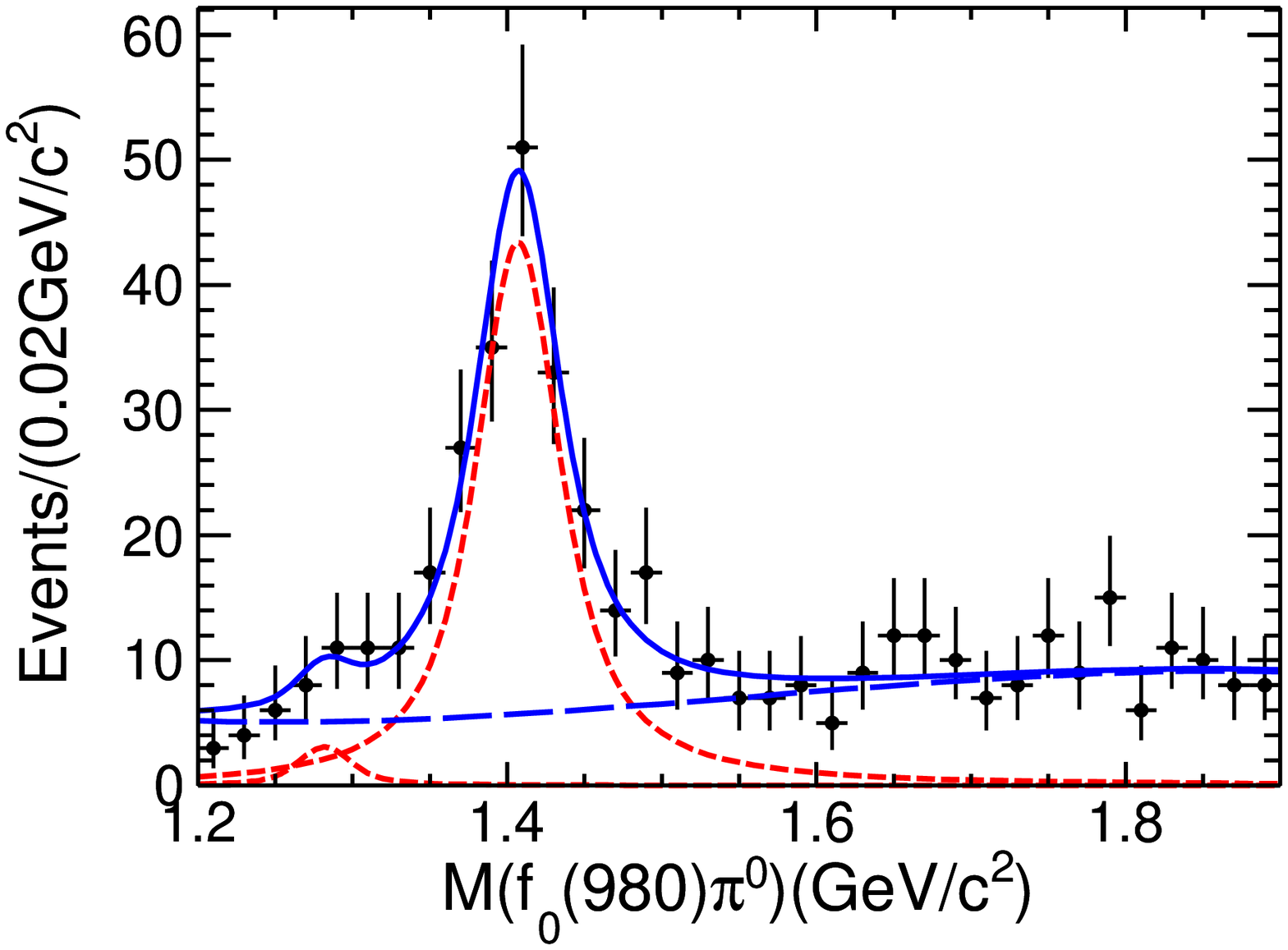,width=4.cm, angle=0}
              \put(-20,70){(b)}}
\centerline{
   \psfig{file=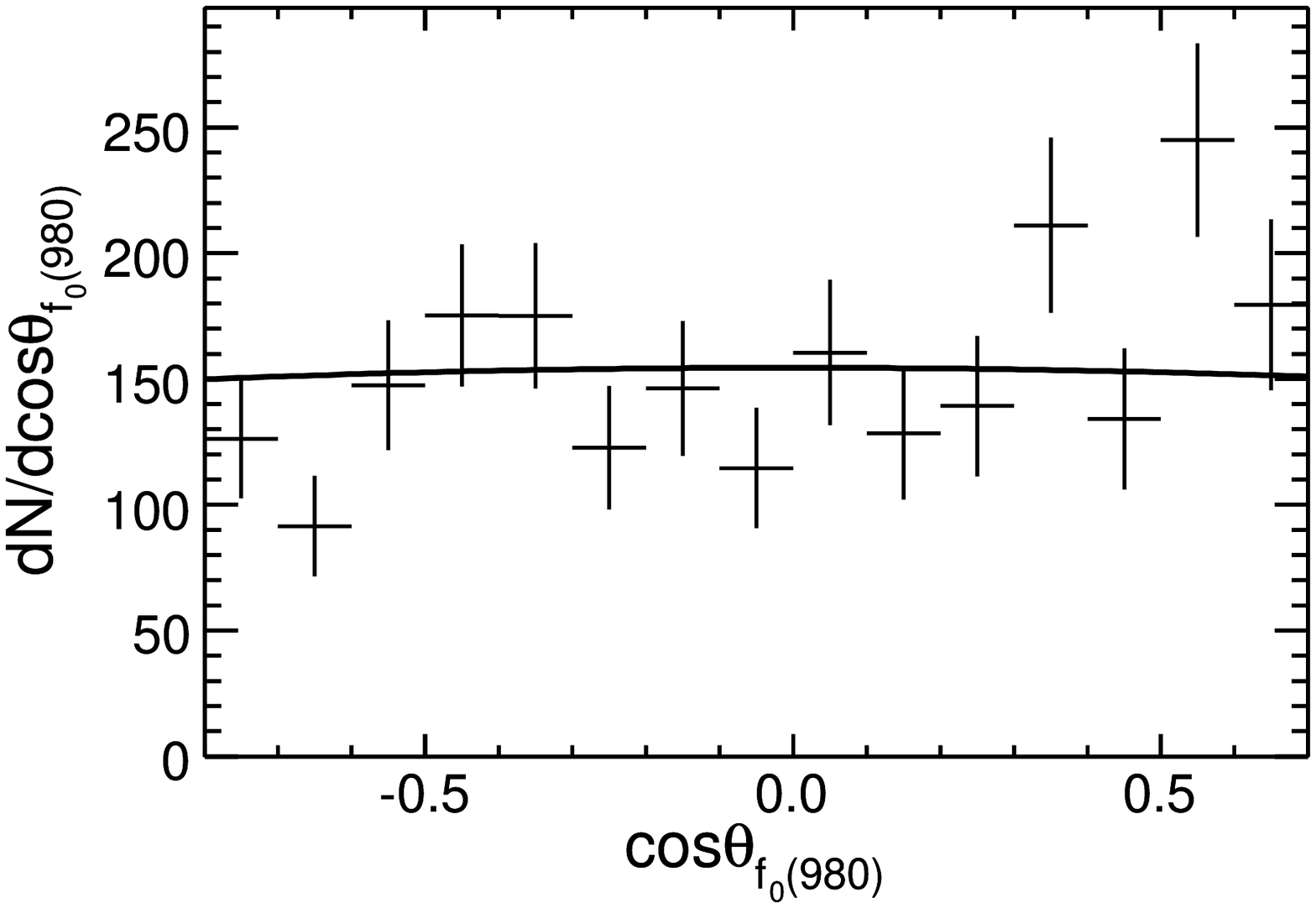,width=4.cm, angle=0}
              \put(-90,20){(c)}
   \psfig{file=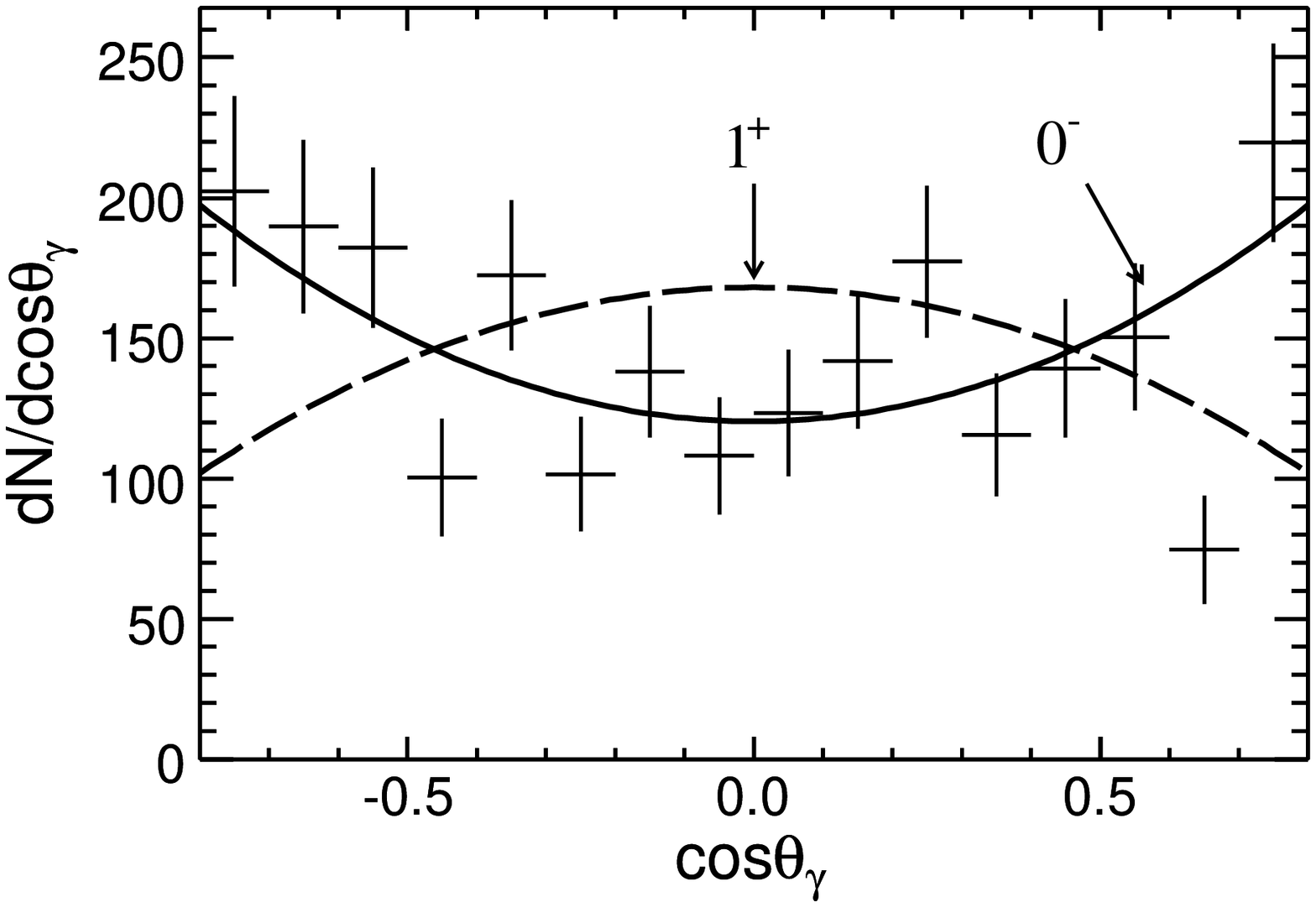,width=4.cm, angle=0}
              \put(-90,20){(d)}}
\caption{\label{fig:fit_result_both} Results of the fit to (a) the
  $f_{0}(980)(\pi^+\pi^-)\pi^0$ and (b) $f_{0}(980)(\pi^0\pi^0)\pi^0$
  invariant mass spectra. The solid curve is the result of the fit
  described in the text. The dotted curve is the $f_{1}(1285)/\eta(1295)$ and
  $\eta(1405)$ signal. The dashed curves denote the background
  polynomial. Angular distributions of the signal include efficiency
  corrections. (c) The $\cos\theta _{f_{0}(980)}$ distribution. The
  fitting result of $\cos\theta _{f_{0}(980)}$ is $|\alpha
  |^2=2.10\pm0.26$. (d) The $\cos\theta _{\gamma}$ distribution. The
  solid line is the prediction for the $J^{P}=0^{-}$ hypothesis, and
  the dashed line is the prediction for the $J^{P}=1^{+}$ hypothesis
  with $|\alpha |^2=2.10$.}
\end{figure}

Figs.~\ref{fig:fit_result_both} (a) and (b) show the
$\pi^{+}\pi^{-}\pi^{0}$ and $\pi^{0}\pi^{0}\pi^{0}$ mass spectra where
$\pi^{+}\pi^{-}(\pi^{0}\pi^{0})$ is in the $f_0(980)$ mass region
(0.94~GeV/c$^2<M_{\pi^{+}\pi^{-}(\pi^{0}\pi^{0})}<1.04$~GeV/c$^2$).
In addition to the $\eta(1405)$, there is an enhancement at
1.3~GeV/$c^2$. A fit is performed to the $\pi^{+}\pi^{-}\pi^{0}$ mass
spectrum. The two peaks are parameterized by efficiency-corrected
Breit-Wigner functions convolved with a Gaussian resolution function.
The mass and width of the small enhancement are fixed to PDG values of
$f_{1}(1285)$~\cite{PDG}. The background is described by a third-order
Chebychev polynomial with shape parameters determined from a
simultaneous fit to the $\pi^{+}\pi^{-}\pi^{0}$ mass spectrum while
the $\pi^{+}\pi^{-}$ invariant mass is found in the $f_{0}(980)$
sidebands. The normalization of each component is allowed to
float. The masses, widths, number of events, efficiencies, and the product branching
ratios of the $\eta(1405)$ and the possible $f_{1}(1285)/\eta(1295)$ contribution
are listed in Table~\ref{table-fit-result}. The statistical
significance of the $\eta(1405)$ is determined by the change of the
fit likelihood -2ln$L$ obtained from the fits with and without the
assumption of the $\eta(1405)$ and is found to be well above
10$\sigma$. The significance of the potential $f_{1}(1285)/\eta(1295)$
contribution is determined to be 3.7$\sigma$. A fit to the
$\pi^{0}\pi^{0}\pi^{0}$ mass spectrum is performed in a similar
fashion, shown in Fig.~\ref{fig:fit_result_both} (b). The significance
of the $\eta(1405)$ is determined to be larger than 10$\sigma$. For a
possible $f_{1}(1285)/\eta(1295)$ contribution, the significance is only
1.2$\sigma$, and we derive an upper limit on the branching ratio at
the 90\% C.L. using the Bayesian method.

An angular-distribution analysis is performed with the selected
$J/\psi\to\gamma \eta(1405)\to\gamma
f_0(980)\pi^0\to\gamma\pi^+\pi^-\pi^0$ events. Backgrounds are
subtracted using the $f_0(980)$ sideband events. For radiative $J/\psi$
decays to a $J^{P}=0^{-}$ meson, the polar angle $\theta_\gamma$ of
the photon in the $J/\psi$ center-of-mass system should be distributed
according to $1+\cos^2\theta_\gamma$. In the case of a $J^{P}=1^{+}$
meson, the distributions $\frac{{d\sigma }}{{d\cos \theta _{\gamma} }}
\sim 1 + 2|\alpha |^2 + (1 - 2|\alpha |^2 )\cos ^2 \theta _{\gamma}$
and $\frac{{d\sigma }}{{d\cos \theta _{f_{0}(980)} }} \sim 2 +
(|\alpha |^2 - 2)\sin ^2 \theta _{f_{0}(980)}$ are expected, where
$\theta _{f_{0}(980)}$ is the polar angle of $f_{0}(980)$ in the
helicity frame of $\eta(1405)$, $\alpha$ is the ratio of helicity 1 to
helicity 0. For the $J^{P}=1^{+}$ assumption, $|\alpha |^2$ is
determined to be $2.10\pm0.26$ from a fit to $\cos\theta_{f_{0}(980)}$
(Fig.~\ref{fig:fit_result_both} (c)). The fitting $\chi^{2}/n.d.f.$ of
Fig.~\ref{fig:fit_result_both} (d) with $|\alpha |^2=2.10$ is
59.4/15. For the $J^{P}=0^{-}$ assumption, the fitting
$\chi^{2}/n.d.f.$ of Fig.~\ref{fig:fit_result_both} (d) is 38.4/15. Comparing the probability of $1^{+}$ hypothesis to $0^{-}$ hypothesis, the ratio of the probabilities is $4.1\times10^{-4}$. The
fitting results favor the $J^{P}=0^{-}$ assignment of the
$\eta(1405)$.

\begin{table*}[hbtp]
  \centering
  \footnotesize
\vspace{-0.0cm} \caption{Summary of measurements of the masses, widths, number of events, the MC efficiencies($\epsilon$) and the product branching ratios of $Br(J/\psi\rightarrow\gamma X)$$\times$$Br(X\rightarrow\pi^{0}f_{0}(980))$$\times$$Br(f_{0}(980)\rightarrow\pi\pi)$ and the decay branching ratios of $Br( Y\rightarrow 3\pi)$ for the $\pi^+\pi^-\pi^0$ and $\pi^0\pi^0\pi^0$ channels, where $X$ represents $\eta(1405)$ and the possible $f_{1}(1285)/\eta(1295)$ contribution, $Y$ represents $\eta'$. Here for the branching ratios, the first errors are statistical and the second ones are systematic.}
\vspace{0.3cm}
\begin{tabular*}{140mm}{l@{\extracolsep{\fill}}ccccc}
\hline
\hline
Resonance              &M(\MeV)         &$\Gamma$(MeV/c$^2$)   &$N_{event}$          &$\epsilon$(\%)  &Branching ratios\\
\hline
$\eta(1405)$($\pi^+\pi^-\pi^0$)        &$1409.0\pm1.7$  &$48.3\pm5.2$   &$743\pm56$ &$22.20\pm0.21$ &$(1.50\pm0.11\pm0.11)\times10^{-5}$\\
$\eta(1405)$($\pi^0\pi^0\pi^0$)        &$1407.0\pm3.5$  &$55.0\pm11.0$  &$198\pm23$ &$12.83\pm0.11$ &$(7.10\pm0.82\pm0.72)\times10^{-6}$\\

$f_{1}(1285)/\eta(1295)$($\pi^+\pi^-\pi^0$)       &fixed           &fixed        &$60\pm18$  &$26.99\pm0.23$ &$(9.99\pm3.00\pm1.03)\times10^{-7}$\\
$f_{1}(1285)/\eta(1295)$($\pi^0\pi^0\pi^0$)       &fixed           &fixed        &$23$   &$16.75\pm0.13$ &$<7.11\times10^{-7}$\\

$\eta'$($\pi^+\pi^-\pi^0$)             &fixed           &fixed        &$1014\pm39$   &$22.52\pm0.19$ &$(3.83\pm0.15\pm0.39)\times10^{-3}$\\
$\eta'$($\pi^0\pi^0\pi^0$)             &fixed           &fixed        &$309\pm19$   &$7.57\pm0.08$ &$(3.56\pm0.22\pm0.34)\times10^{-3}$\\
\hline
\hline
\end{tabular*}%
        \label{table-fit-result}
\end{table*}

\begin{figure}[htbp]
   \centerline{
   \psfig{file=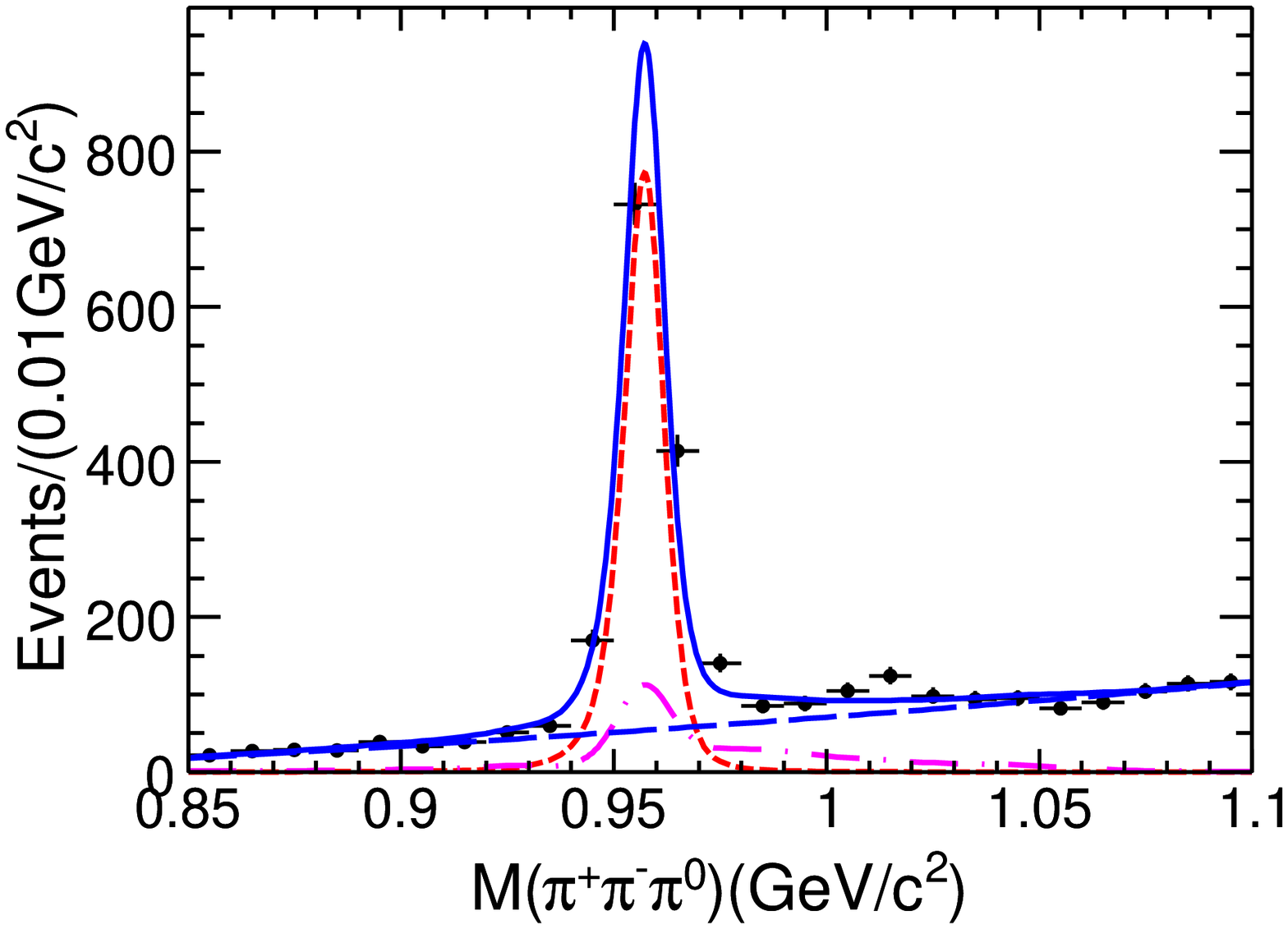,width=4.cm, angle=0}
              \put(-20,70){(a)}
   \psfig{file=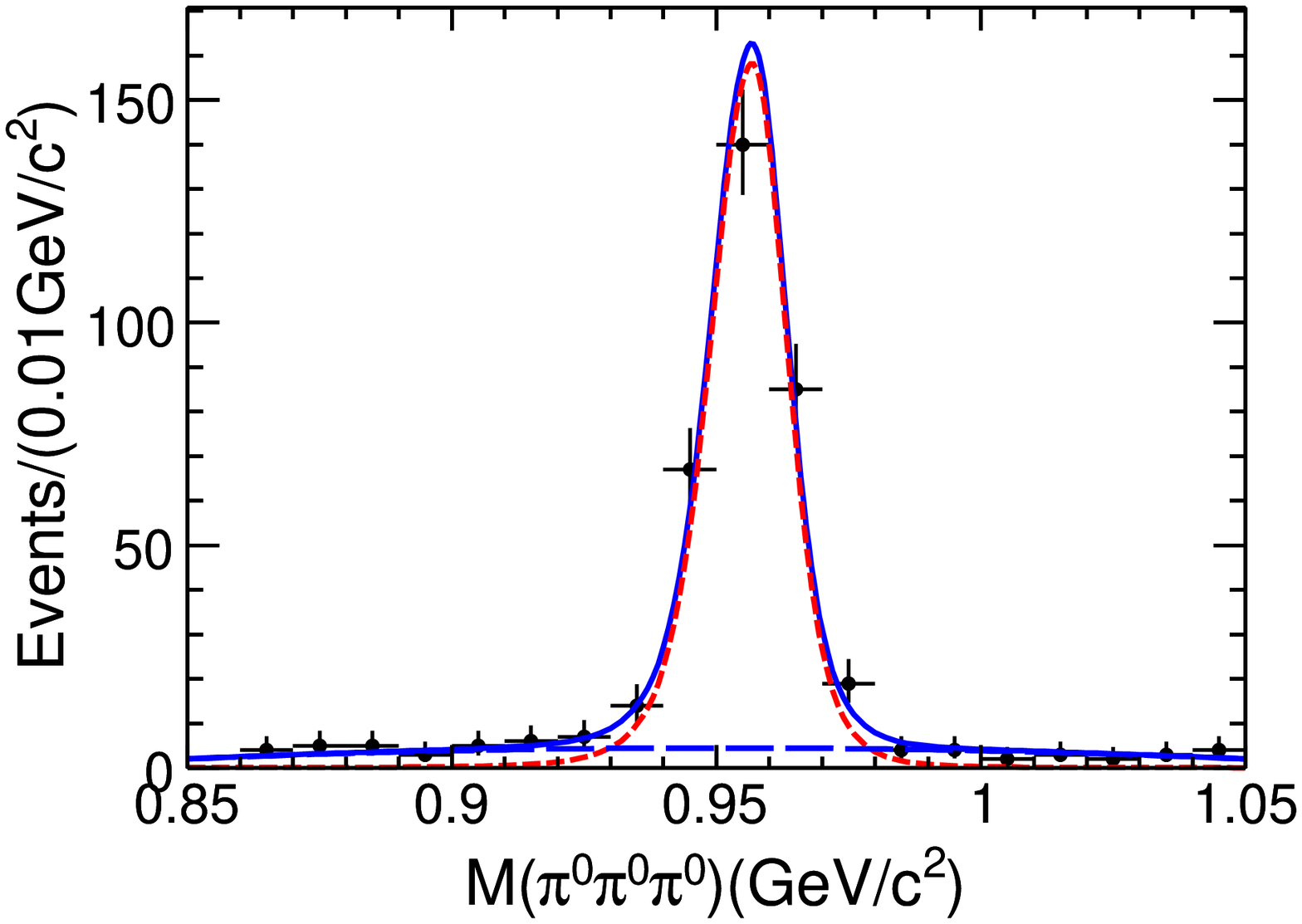,width=4.cm, angle=0}
              \put(-20,70){(b)}}
    \caption{\label{fig:fit_result_etap} Results of the fit to (a) the
    $\pi^+\pi^-\pi^0$ and (b) $3\pi^0$ invariant mass spectra. The solid curve is the result of the fit described in the text. The dotted curve is the $\eta'$ signal. The dashed curves denote the background polynomial. The dash-dotted curve in (a) describes the peaking background.}
\end{figure}

Figs.~\ref{fig:fit_result_etap} (a) and (b) show the
$\pi^{+}\pi^{-}\pi^{0}$ and $\pi^{0}\pi^{0}\pi^{0}$ mass spectra in
the $\eta'$ region. A fit is performed to the $\pi^{+}\pi^{-}\pi^{0}$
mass spectrum, shown in Fig.~\ref{fig:fit_result_etap} (a). The shape
of the $\eta'$ is obtained from MC simulation and the mass and width
of the $\eta'$ are fixed to its PDG values~\cite{PDG}. The shape of
the peaking backgrounds
($J/\psi\to\gamma\eta'\to\gamma\gamma\rho^{0}(\pi^+\pi^-)$ and
$J/\psi\to\gamma\eta'\to\gamma\gamma\omega(\pi^+\pi^-\pi^0)$) are from
exclusive MC samples with predicted background levels of 361$\pm$32
and 32$\pm$6 events, normalized by branching ratios in the
PDG~\cite{PDG} and fixed in the fit. The error on the number of events
is estimated by changing the normalization by one standard deviation
from the PDG value. A second-order Chebychev polynomial is used to
describe the sum of other non-peaking backgrounds. For
$\eta'\rightarrow\pi^0\pi^0\pi^0$,
$\chi^2(\gamma\pi^0\pi^0\pi^0)<\chi^2(\gamma\eta\pi^0\pi^0)$ and
$|M_{\gamma\gamma}-m_{\eta}|>0.03$ GeV/$c^2$ are additionally required
to remove background events from $\eta'\rightarrow\eta\pi^0\pi^0$. A
fit to the $\pi^{0}\pi^{0}\pi^{0}$ mass spectrum is performed as shown
in Fig.~\ref{fig:fit_result_etap} (b). The shapes of the $\eta'$ and
non-peaking backgrounds are described analogously to the charged
mode. The efficiencies and the product branching ratios for the
$\eta'$ obtained from the fit are also listed in
Table~\ref{table-fit-result}. From our measurement and the world
average values for branching ratios of
$\eta'\to\eta\pi\pi$~\cite{PDG}, we determine $r_{\pm} = (8.87\pm
0.98)\times10^{-3}$ and $r_{0} = (16.41\pm 1.94)\times10^{-3}$.

The systematic uncertainties on the signal yield arise from fit
ranges, signal shapes and background estimation. In detail, for the
$\eta(1405)$ signal, the signal shape is given by a Breit-Wigner
function with floating mass and width. Its uncertainties are estimated
by fixing the mass and width of the $\eta(1405)$ to the world average
values. For the possible $f_{1}(1285)/\eta(1295)$ contribution and $\eta'$
signal, uncertainties are estimated by changing their mass and width
by one standard deviation from the PDG values. The uncertainty due to
the assumed background shape for the $\eta(1405)$ and the
$f_{1}(1285)/\eta(1295)$ has been estimated using different $f_{0}(980)$
sidebands; while that of the $\eta'$ is studied using different order
polynomials. For $\eta'\rightarrow\pi^{+}\pi^{-}\pi^{0}$, the
uncertainty of the peaking background is estimated by using the shape
from the $\pi^{0}$ sidebands instead of using the shape from exclusive
MC samples. The systematic errors on the branching ratio measurements
are also subject to systematic uncertainties in the number of $J/\psi$
events~\cite{ref:jpsitotnumber}, the intermediate branching
ratios~\cite{PDG}, the data-MC difference in the $\pi$ tracking
efficiency, the photon detection efficiency, particle identification,
the kinematic fit and the $\pi^{0}$ selection. Combined in quadrature
with the uncertainty from the mass spectrum fitting, the systematic
errors on the product branching ratios of the $\eta(1405)$,
$f_{1}(1285)/\eta(1295)$ and $\eta'$ are summarized in
Table~\ref{table-fit-result}.

The $\eta(1405)\to f_{0}(980)\pi^0$ signal could arise via $\eta(1405)\to a_{0}^{0}(980)\pi^0$ and $a_0^0(980)-f_0(980)$ mixing. Using the branching ratio of
$J/\psi\rightarrow\gamma\eta(1405)\rightarrow\gamma\eta\pi^{+}\pi^{-}$
and the largest PDG value of $\Gamma(\eta(1405)\to
a_{0}(980)\pi)$/$\Gamma(\eta(1405)\to\eta\pi\pi)$, $Br(J/\psi\rightarrow\gamma\eta(1405)\rightarrow\gamma
a_{0}^{0}(980)\pi^{0}\rightarrow\gamma\pi^{0}\eta\pi^{0})=
(8.40\pm1.75)\times10^{-5}$. The $a_0^0(980)-f_0(980)$ mixing
intensity ($\xi_{af}=Br(\chi_{c1}\rightarrow
f_{0}(980)\pi^0\rightarrow\pi^+\pi^-\pi^0)/Br(\chi_{c1}\rightarrow
a_{0}^{0}(980)\pi^0\rightarrow\eta\pi^0\pi^0)$) measured at BESIII is less than 1\%
(90\% C.L.)~\cite{ref:wangyd_a0f0}.  The branching ratio of
$J/\psi\to\gamma\eta(1405)\to\gamma a_{0}^{0}(980)\pi^0\to\gamma
f_{0}(980)\pi^0\to\gamma\pi^+\pi^-\pi^0$ is thus expected to be less
than $(8.40\pm1.75)\times 10^{-7}$, which is much smaller than the
result that we measure. Therefore, $a_0^{0}(980)-f_0(980)$ mixing
alone can not explain the observed branching ratio of $\eta(1405)\to
f_{0}(980)\pi^0$.

To interpret the anomalously narrow width of $f_{0}(980)$ and large isospin violation observed in this channel, J.J. Wu et al. recently propose that a novel scenario of triangle singularity would play a crucial role in this process~\cite{ref:wujiajun}. Further theoretical and experimental studies are needed for a better understanding of the underlying dynamics.

In summary, we have studied $J/\psi\rightarrow\gamma
3\pi$ decays. The isospin violating decay $\eta(1405)\rightarrow
f_0(980)\pi^0$ is observed for the first time with a statistical
significance larger than $10\sigma$ in both the charged and neutral
modes.
According to our measurement, the ratio of $Br(\eta(1405)\rightarrow
f_{0}(980)\pi^{0} \rightarrow\pi^{+}\pi^{-}\pi^{0})$ to
$Br(\eta(1405)\rightarrow a_{0}^{0}(980)\pi^{0} \rightarrow
\eta\pi^{0}\pi^{0})$ is $(17.9\pm4.2)\%$~\cite{PDG, CBeta1440}, which
is one order of magnitude larger than the $a_0^{0}(980)-f_0(980)$
mixing intensity (less than 1\%) determined at BESIII previously
~\cite{ref:wangyd_a0f0}. The measured width of the $f_{0}(980)$ is anomalously narrower than the world average.
There is also evidence for an enhancement at around
1.3~GeV/$c^2$ (potentially from the $f_{1}(1285)/\eta(1295)$) seen with a
significance of 3.7$\sigma$ in the charged mode and 1.2$\sigma$ in the
neutral mode.
For
the decay $\eta'\rightarrow\pi^+\pi^-\pi^0$, the branching ratio that
we measure is consistent with the CLEO-c
measurement~\cite{ref:CLEO_g3pi}, and the precision is improved by a
factor of four. For the decay $\eta'\rightarrow 3\pi^0$, it is two
times larger than the world average value~\cite{PDG}. Using our new
measurement of the decay rates for $\eta'\to 3\pi$, the values of $r_{\pm}$ and $r_{0}$ are more than four standard deviations away from both $\pi^0$-$\eta$ mixing prediction and the chiral unitary framework prediction~\cite{ref:etaptheo2}.

The BESIII collaboration thanks the staff of BEPCII and the computing center for their hard efforts. Useful discussions with K.T. Chao, S.L. Zhu, and J.J. Wu are acknowledged. This work is supported in part by the Ministry of Science and Technology of China under Contract No. 2009CB825200; National Natural Science Foundation of China (NSFC) under Contracts Nos. 10625524, 10821063, 10825524, 10835001, 10935007, 11125525; the Chinese Academy of Sciences (CAS) Large-Scale Scientific Facility Program; CAS under Contracts Nos. KJCX2-YW-N29, KJCX2-YW-N45; 100 Talents Program of CAS; Istituto Nazionale di Fisica Nucleare, Italy; Siberian Branch of Russian Academy of Science, joint project No 32 with CAS; U. S. Department of Energy under Contracts Nos. DE-FG02-04ER41291, DE-FG02-91ER40682, DE-FG02-94ER40823; University of Groningen (RuG) and the Helmholtzzentrum fuer Schwerionenforschung GmbH (GSI), Darmstadt; WCU Program of National Research Foundation of Korea under Contract No. R32-2008-000-10155-0.


\begin{thebibliography}{99}
%
\bibitem{ref:first1440}P.H. Baillon {\it et al.}, Nuovo Cimento {\bf 50A}, 393 (1967).


\bibitem{ref:glue1440}M. Acciarri {\it et al.} (L3 Collaboration), Phys. Lett. B {\bf 501}, 1 (2001).
\bibitem{ref:glueFad}L. Faddeev, A.J. Niemi and U. Wiedner, Phys. Rev. D {\bf 70}, 114033 (2004).

\bibitem{ref:latticeQCD}G. S. Bali {\it et al.}, Phys. Lett. B {\bf 309}, 378 (1993).

\bibitem{ref:status1440}A. Masoni, C. Cicalo, and G.L. Usai, J. Phys. {\bf G32}, R293 (2006).

\bibitem{ref:a0f0-1}R. L. Jaffe, Phys. Rev. D {\bf 15}, 267 (1977).

\bibitem{ref:a0f0-2}N. N. Achasov and V. N. Ivanchenko, Nucl. Phys. B {\bf 315}, 465 (1989).

\bibitem{ref:a0f0-3}N. N. Achasov and V.V. Gubin, Phys. Rev. D {\bf 56}, 4084 (1997).

\bibitem{ref:a0f0-4}J. D. Weinstein and N. Isgur, Phys. Rev. D {\bf 27}, 588 (1983).

\bibitem{ref:a0f0-5}J. D. Weinstein and N. Isgur, Phys. Rev. D {\bf 41}, 2236 (1990).

\bibitem{ref:a0f0-6}S. Ishida {\it et al.}, in Proceedings of the 6th International
Conference on Hadron Spectroscopy, Manchester, United
Kingdom, 1995 (World Scientific, Singapore, 1995), p. 454.

\bibitem{Achasov:1979xc}
  N.~N.~Achasov, S.~A.~Devyanin and G.~N.~Shestakov,
  Phys.\ Lett.\  B {\bf 88}, 367 (1979).


\bibitem{Hanhart:2007bd}
  C.~Hanhart, B.~Kubis and J.~R.~Pelaez,
  Phys.\ Rev.\  D {\bf 76}, 074028 (2007).


\bibitem{Achasov:2002hg}
  N.~N.~Achasov and A.~V.~Kiselev,
  Phys.\ Lett.\  B {\bf 534}, 83 (2002).


\bibitem{Kerbikov:2000pu}
  B.~Kerbikov and F.~Tabakin,
  Phys.\ Rev.\  C {\bf 62}, 064601 (2000).


\bibitem{Achasov:2003se}
  N.~N.~Achasov and G.~N.~Shestakov,
  Phys.\ Rev.\ Lett.\  {\bf 92}, 182001 (2004).


\bibitem{Close:2000ah}
  F.~E.~Close and A.~Kirk,
  Phys.\ Lett.\  B {\bf 489}, 24 (2000).


%
%

\bibitem{ref:etaptheo1}D.J. Gross, S.B. Treiman, and F. Wilczek, Phys. Rev. D {\bf 19}, 2188(1979).

\bibitem{ref:etaptheo2}B. Borasoy, U.-G. Meissner, and R. Nissler, Phys. Lett. B {\bf 643}, 41(2006).

%

\bibitem{ref:jpsitotnumber} M.~Ablikim {\it et al.} (BESIII Collaboration), Phys. Rev. D {\bf 83}, 012003 (2011).

\bibitem{ref:bes3nim} M.~Ablikim {\it et al.} (BESIII Collaboration), Nucl. Instrum. Methods Phys. Res., Sect. A {\bf 614}, 345 (2010).

\bibitem{bepc2_design} J. Z. Bai {\it et al.} (BES Collaboration), Nucl. Instrum. Methods Phys. Res. A {\bf 458}, 627 (2001).

\bibitem{ref:geant4} S. Agostinelli {\it et al.} (\textsc{geant}{\footnotesize
4} Collaboration), Nucl. Instrum. Methods Phys. Res. A {\bf 506}, 250 (2003).

\bibitem{ref:geant4_2} J. Allison {\it et al.}, IEEE Trans. Nucl. Sci. {\bf 53}, 270 (2006).

\bibitem{ref:bes3gen} R.~G.~Ping, Chinese Phys. C {\bf 32}, 599 (2008).
%
\bibitem{PDG}K. Nakamura {\it et al.} (Particle Data Group), J. Phys. G {\bf 37}, 075021 (2010).

\bibitem{CBeta1440}C.Amsler {\it et al}. (Crystal Barrel Collaboration), Phys. Lett. B {\bf 358}, 389 (1995).

\bibitem{ref:wangyd_a0f0} M.~Ablikim {\it et al.} (BESIII Collaboration), Phys. Rev. D {\bf 83}, 032003 (2011).

\bibitem{ref:wujiajun} J.J. Wu, X.H. Liu, Q. Zhao, B.S. Zou, Phys. Rev. Lett. {\bf 108}, 081803 (2012).

\bibitem{ref:CLEO_g3pi} P. Naik {\it et al.} (CLEO Collaboration), Phys. Rev. Lett. {\bf 102}, 061801 (2009).
%


\end{thebibliography}
\end{document}